\documentstyle[prl,aps,epsfig,multicol]{revtex}

\def\Journal#1#2#3#4{{#1}{\bf #2}, #3 (#4)}

\def\NPA{{Nucl. Phys.}~{\bf A}}

\def\PRL{Phys. Rev. Lett.\ }
\def\PRD{{Phys. Rev.}~{\bf D}}
\def\PRC{{Phys. Rev.}~{\bf C}}
\def\ZPC{{Z. Phys.}~{\bf C}}

\begin{document}
\draft

\title{Measurement of the mid-rapidity transverse energy distribution \\
      from $\sqrt{s_{_{NN}}}=130$~GeV Au+Au collisions at RHIC}

%
%

\author{
K.~Adcox,$^{40}$
S.{\,}S.~Adler,$^{3}$
N.{\,}N.~Ajitanand,$^{27}$
Y.~Akiba,$^{14}$
J.~Alexander,$^{27}$
L.~Aphecetche,$^{34}$
Y.~Arai,$^{14}$
S.{\,}
% ...
\\(PHENIX Collaboration)\\
}
\address{
% ...
$^{3}$Brookhaven National Laboratory, Upton, NY 11973-5000, USA\\
% ...
$^{14}$KEK, High Energy Accelerator Research Organization, Tsukuba-shi, Ibaraki-ken 305-0801, Japan\\
% ...
$^{27}$Chemistry Department, State University of New York - Stony Brook, Stony Brook, NY 11794, USA\\
% ...
$^{34}$SUBATECH (Ecole des Mines de Nantes, IN2P3/CNRS, Universite de Nantes) BP 20722 - 44307, Nantes-cedex 3, France\\
% ...
$^{40}$Vanderbilt University, Nashville, TN 37235, USA\\
% ...
}

\date{\today}        

\maketitle

\begin{abstract}
The first measurement of energy produced transverse to the beam 
direction at RHIC is presented. 
The mid-rapidity transverse 
energy density per participating nucleon rises steadily with the number of
participants, closely paralleling the rise in charged-particle density, such 
that $\langle E_T\rangle/\langle N_{ch}\rangle$ remains relatively constant 
as a function of centrality. 
The energy density calculated via Bjorken's 
prescription for the 2\% most central Au+Au collisions at
$\sqrt{s_{_{NN}}}=130$~GeV is at least $\epsilon_{Bj}=4.6$~GeV/fm$^3$, 
which is a factor of 1.6 larger than found at 
$\sqrt{s_{_{NN}}}=17.2$~GeV (Pb+Pb at CERN). 
\end{abstract}


\begin{multicols}{2}
\narrowtext


The PHENIX detector~\cite{PHENIX1} at RHIC, the Relativistic Heavy Ion
Collider at Brookhaven National Laboratory, is designed to measure the
properties of nuclear matter at the highest temperatures and energy
densities. 
For example a transition to a quark-gluon plasma has been predicted 
for energy densities on the order of a few GeV/fm$^3$~\cite{QMproc}. 
The spatial energy density ($\epsilon$) in a relativistic collision can be 
estimated (following Bjorken~\cite{Bj83}) 
by measuring the 
transverse energy density in rapidity, $dE_T/dy$, which is effectively
the co-moving energy density in a longitudinal expansion:
\begin{equation}
\epsilon_{Bj}={d E_T \over dy} {1\over \tau_0\,\pi R^2}
\label{eq:9} \end{equation} 
where $\tau_0$, the formation time, is usually taken as ${1}$ fm/c,
and $\pi R^2$ is the effective area of the collision.  The transverse
energy ($E_T$) is a multiparticle variable defined as: 
\begin{equation} 
E_T=\sum_{i} E_i \sin\theta_i , \ 
dE_T (\eta)/d\eta=\sin\theta (\eta)\,dE (\eta)/d\eta , 
\label{eq:ETdef}
\end{equation} 
where $\theta$ is the polar angle, 
$\eta=-\ln \tan\theta/2$ is the pseudorapidity, $E_i$ is by
convention taken as the kinetic energy for nucleons and the total energy
for all other particles~\cite{NA34}, and the sum is taken over all
particles emitted into a fixed solid angle for each event. 
$E_T$
measurements, even in limited apertures at mid-rapidity, provide
excellent characterization of the nuclear geometry of a reaction on an
event-by-event basis and are sensitive to the underlying reaction
dynamics~\cite{QMproc}.


During the RHIC run in the summer of 2000, 
PHENIX accumulated close to 5 million 
interaction triggers for Au+Au collisions at $\sqrt{s_{_{NN}}}=130$
GeV using Zero Degree Calorimeters (ZDC) and Beam-Beam Counters (BBC)
as triggering devices.  The events were selected with a requirement on the 
collision vertex position along the beam axis, $|z|\leq 20$ cm, 
as in the recent PHENIX publication on mid-rapidity multiplicity
distributions~\cite{PXpub2001-1}, where further details are given.

	The present measurement uses a section of the electromagnetic 
calorimeter (EMCal) from the PHENIX central-spectrometer, with front face 
5.1~m from the beam axis. 
This section is part of a sampling calorimeter, custom developed and built 
for PHENIX~\cite{rf:calor}, composed 
of alternating Pb and scintillator tiles (PbSc) with readout of individual 
towers, $5.54\times5.54$ cm$^2$ in cross section, via wavelength shifting 
(WLS) fibers in a ``shashlik'' geometry.  
The depth of the PbSc calorimeter is $18$ radiation lengths ($X_0$) which 
corresponds to $0.85$ interaction lengths. 
The PbSc calorimeter has an energy resolution of 
8.2\%/{$\sqrt {E({\rm GeV})}$} $\oplus 1.9$\% for test beam electrons, 
with measured response proportional to 
incident electron energy to within $\pm 2$\% over the range 
$0.3 \leq E_e \leq 40.0$~GeV~\cite{rf:calor}.

\begin{figure}

\centerline{\epsfig{file=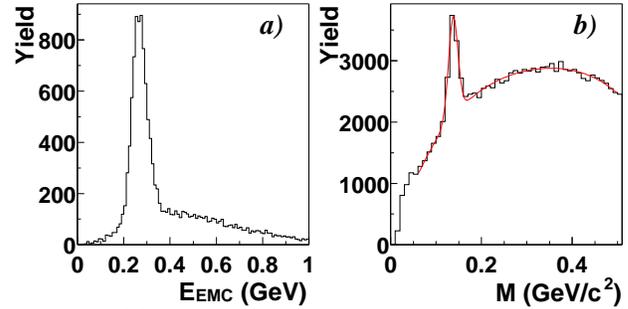,width=8.5cm}}
\caption[]{(a) The distribution of EMCal clusters 
corresponding to 1 GeV/c charged tracks (mostly pions) from Au+Au collisions. 
(b) The reconstructed $\pi^0$ mass from pairs of EMCal clusters with total 
energy $>2$~GeV.}
\label{fig:1} 
\end{figure}

	During construction, the calibration of the 
calorimeter was set by simultaneously recording the response to laser 
excitation and to cosmic-ray muons penetrating transversely to the tower axis. 
The calibration was maintained in-situ during the run by 
monitoring relativistic charged particles from Au+Au collisions. The 
absolute energy scale was 
determined by test-beam measurements normalized to electrons with known 
energy.  A final adjustment of the absolute energy scale was performed 
using in-situ identified electrons ($p>500$
MeV/c) by shifting the originally measured energy/momentum $(E/p)$ peak  
from 1.02$\pm$0.01 to 1.00. The accuracy of the absolute energy scale was 
cross-checked in-situ against both the minimum ionizing peak (MIP) of 
charged particles penetrating along 
the tower axis and the mass of the $\pi^0$.
The corrected energy distribution of EMCal 
clusters from $1.0\pm0.1$~GeV/c charged tracks (mostly pions)
measured in the Drift Chamber~\cite{PHENIX1} 
exhibits a clear MIP (Fig.~\ref{fig:1}a),
as well as energy due to nuclear interactions in the material of
the EMCal. The MIP position is in agreement within 2\% to the
value obtained in the test beam (270 MeV).  The mass of 
the $\pi^0$, reconstructed from pairs of EMCal clusters (assumed to be 
photons~\cite{photon}) of total energy 
greater than 2 GeV (Fig.~\ref{fig:1}b), is within 1.5\% of the
published value. This sets the systematic error of the absolute energy scale 
at less than $1.5$\%. 

  The data sample for the present $E_T$ measurement is taken from the same 
runs used in our multiplicity measurement~\cite{PXpub2001-1} 
(no magnetic field), and comprises about 140,000 events from the BBC 
trigger which detects [$92\pm2(syst)$]\% of the nuclear
interaction cross section of 7.2b with a background contamination of
[$1\pm1(syst)$]\%~\cite{PXpub2001-1}.  The transverse energy was 
measured using the PbSc EMCal in a fiducial aperture $|\eta|\leq 0.38$ in
pseudorapidity and $\Delta\phi=44.4^\circ$ in azimuth. 
$E_T$ was computed for each event (Eq.~\ref{eq:ETdef}) using clusters of 
energy greater than 20 MeV, composed of adjacent towers with 
deposited energy of more than 3 MeV.  The angle $\theta_i$ is computed from 
the centroid of the cluster of energy $E_i$ assuming a particle originating 
from the event vertex. 

\begin{figure}

\centerline{\epsfig{file=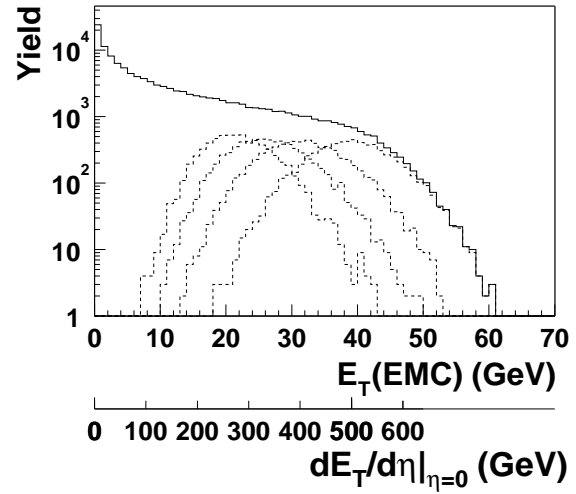,width=8.5cm}}
\caption[]{The raw ${E_T}_{\rm EMC}$ distribution measured in the 
$\Delta\phi=44.4^\circ$ azimuthal and $|\eta|\leq 0.38$ polar angle fiducial  
acceptance for Au+Au at $\sqrt{s_{_{NN}}}=130$~GeV (upper scale) and 
total hadronic $dE_T/d\eta|_{\eta=0}$ (lower scale), see text. The solid line 
is the minimum bias distribution with the BBC trigger; the dashed lines 
correspond to the distributions for the 4 most central bins 
in Table I.}
\label{fig:2} 
\end{figure}

The raw spectrum of measured transverse energy, ${E_T}_{\rm EMC}$, in the 
fiducial aperture of the PHENIX EMCal for Au+Au collisions at 
$\sqrt{s_{_{NN}}}=130$~GeV is shown in Fig.~\ref{fig:2}, upper scale. 
The lower scale in Fig.~\ref{fig:2} represents a correction of the raw 
${E_T}_{\rm EMC}$ by a factor of 12.8 to correspond to the hadronic 
$dE_T/d\eta|_{\eta=0}$ in the full azimuth. The 12.8 is composed of a  
factor of 10.6 for the fiducial acceptance, a factor of 1.03 for 
disabled calorimeter towers and a  
factor, $k=1.17\pm 0.01$, which is the ratio of the hadronic $E_T$ in the 
fiducial aperture to the measured ${E_T}_{\rm EMC}$. The $k$ factor 
includes the response of the detector to charged and neutral particles emitted 
from the event vertex into the fiducial aperture, and additional corrections 
for energy in-flow from outside the fiducial aperture and for
losses~\cite{explain}.
These factors were calculated 
with a GEANT~\cite{GEANT} based Monte Carlo (MC) simulation of the detector 
using HIJING as the event generator~\cite{HIJING}.

For $E_T$ measurements at mid-rapidity at a collider,   
the EMCal acts as a thin but effective hadronic calorimeter. 
Charged pions with $p_T\leq 0.35$~GeV/c, 
kaons ($p_T\leq0.64$~GeV/c) and protons ($p_T\leq0.94$~GeV/c)---$p_T$ values 
which are near or above the $\langle p_T\rangle$ for all 3 cases---stop 
(i.e. deposit all their kinetic 
energy) in the EMCal. For higher $p_T$ hadrons, 43\% leave the MIP and 57\% 
interact, leaving an average of $\sim 65$\% of their energy. 
The measured ${E_T}_{\rm EMC}$ is $0.79\pm 0.01$ of the total $E_T$ striking 
the EMCal, which is composed roughly of 40\% produced by charged pions, 
40\% by photons (from $\pi^0$ and other decays), and 20\% by 
all other particles (including decay muons).  The particle composition 
and $\langle p_T\rangle$ in HIJING are close to the observed values, and  
furthermore,  the 
$k$ factor is insensitive to reasonable variations (for instance varying the 
momenta of all particles by $\pm 15$\% changes the overall $k$ by less than 
$\pm 2$\%), leading to an estimated systematic uncertainty in $k$ of less than 
${\pm}3$\% due to particle composition and momentum. 

\begin{figure}

\centerline{\epsfig{file=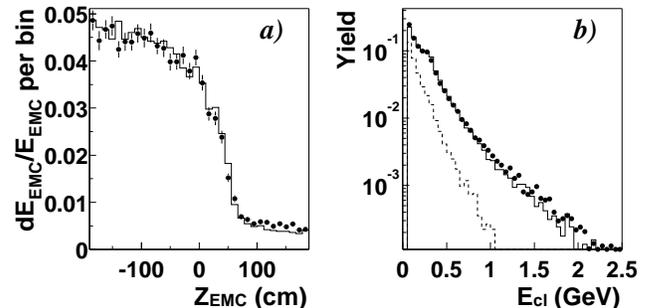,width=8.5cm}}
\caption[]{(a) The fraction of ${E_T}_{\rm EMC}$ in bins of 11.08 cm along 
the EMCal $z_{\rm EMC}$ direction for event vertex near a pole face; 
histogram from MC simulation, solid points from beam data. (b) EMCal cluster 
energy spectrum from HIJING MC (solid line), with in-flow component 
(dotted line), compared to data (solid points).}
\label{fig:3} 
\end{figure}

	The main issues for the MC are the in-flow contribution and losses. 
The losses are due to particles which originate within the aperture but 
whose decay products miss the EMCal (10\%), or whose energy is lost due to 
edge effects (6\%) or clustering (2\%). The in-flow, ($24\pm1$)\% of the 
$E_T$ striking the EMCal, is principally of two types: (1) albedo from the 
magnet poles; (2) particles which originate outside the aperture of the 
calorimeter but whose decay products hit the calorimeter. The in-flow 
component of $k$
was checked by comparing the MC and the
measurements for events with a vertex outside the normal range,  
just at and inside a pole face of the axial central-spectrometer magnet, 
$38\leq z\leq 42$ cm, for which 
the calorimeter aperture is partly shadowed.  The fraction of 
the total energy, $dE_{\rm EMC}/E_{\rm EMC}$, 
in bins of width 2 towers along the $z$ coordinate of the EMCal, 
$z_{\rm EMC}$, is shown in 
Fig.~\ref{fig:3}a.  The HIJING MC simulation agrees with the 
measured data everywhere except in the range $z_{\rm EMC}> 100$ cm, which
is fully shadowed by the pole, where the simulation  shows $\sim$20\%  
less energy than the data. In Fig.~\ref{fig:3}b, the  
distributions of the cluster energy, $E_{cl}$, for the open aperture,
$z_{\rm EMC}< -50$ cm, are shown for both HIJING and the data and are  
in excellent agreement. The in-flow component
of HIJING is also indicated as a dotted line and falls much more sharply
than the total $E_{cl}$ spectrum.  The residual discrepancy of the 
energy in the shadowed region, which contributes roughly 10\% of
the total signal, results in a $\pm (2-3)$\% systematic uncertainty in
$E_T$ due to the uncertainty in the in-flow. Combining this with the 
uncertainty due to particle composition and momentum yields an overall 
factor $k=[1.17\pm 0.01]\pm 4$\%~$(syst)$, which, according to the MC, is 
independent of centrality.

	Returning to Fig.~\ref{fig:2}, the shape of the measured transverse 
energy spectrum shows the 
characteristic form of $E_T$ distributions in limited apertures: a peak
and sharp drop-off at low values of $E_T$ corresponding to peripheral
collisions with grazing impact; a broad, gently sloping plateau at the 
mid-range of impact parameters, dominated by the nuclear geometry; 
and then at higher values of $E_T$, which correspond to the most central
collisions where the nuclei are fully overlapped, a `knee' leading to a 
fall-off which is very steep for large apertures and which becomes 
less steep, the smaller the aperture~\cite{E802000}. It should be emphasized 
that the correction of ${E_T}_{\rm EMC}$ to $dE_T/d\eta|_{\eta=0}$ by a 
single scale factor (predominantly acceptance) is valid up to the knee of 
the distribution, roughly the upper 1 percentile. Above the knee, the
fall-off depends on the aperture and is sensitive to detector effects as 
well as statistical and dynamical fluctuations. 
Thus an actual measurement of 
$dE_T/d\eta|_{\eta=0}$ for $\Delta\eta=1.0$ and full azimuth would have
a sharper fall-off above the knee. With this caveat,  
the uncertainty in the absolute energy scale ($\pm 1.5$\%) and the uncertainty
in $k$ of $\pm 4$\% are combined to yield an overall uncertainty in the 
hadronic $dE_T/d\eta|_{\eta=0}$ of $\pm 4.5$\%~$(syst)$, 
independent of $E_T$, where the statistical error is negligible. 

\begin{figure}

\centerline{\epsfig{file=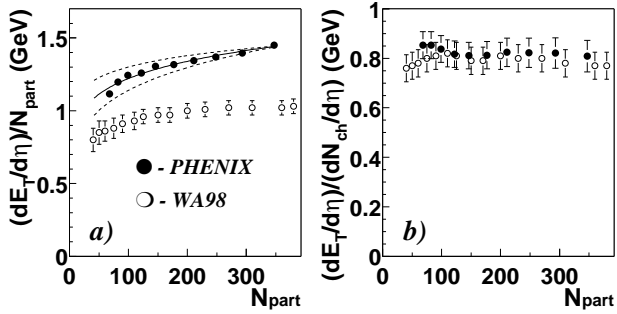,width=8.5cm}}
\caption[]{(a) PHENIX transverse energy density per participant
$dE_T/d\eta|_{\eta=0}/{ N_{\rm part}}$ for 
Au+Au collisions at $\sqrt{s_{_{NN}}}=130$~GeV  as a function of $N_{\rm part}$, 
the number of participants, compared to
data from WA98~\cite{rf:WA98} for Pb+Pb collisions at
$\sqrt{s_{_{NN}}}=17.2$~GeV.  The solid line is the $N_{\rm part}^\alpha$ best fit 
and the dashed lines represent the effect of the $\pm 1\sigma$ 
$N_{\rm part}$-dependent systematic errors for $dE_T/d\eta|_{\eta=0}$ and 
$N_{\rm part}$. There is an additional overall ($N_{\rm part}$-independent) 
systematic uncertainty of $\pm 4.5$\% from $dE_T/d\eta|_{\eta=0}$ and 
$\pm 2.0$\% from $N_{\rm part}$. 
(b) PHENIX $dE_T/d\eta|_{\eta=0}/dN_{ch}/d\eta|_{\eta=0}$ 
versus $N_{\rm part}$, including all systematic errors,  compared to WA98. 
Note that the WA98 data in both (a) 
and (b) have an additional $\pm 20$\% overall systematic error which is 
not shown.} 
\label{fig:4} 
\end{figure}

Mid-rapidity $E_T$ distributions are a standard method of defining 
centrality~\cite{QMproc,E802000,rf:WA98,NA49ET}. Thus, it is 
important to determine for the present data the detailed relationship 
of transverse energy production to $N_{\rm part}$, the number of nucleons 
participating in the collision (participants), 
which in earlier fixed target experiments was deduced 
straightforwardly by measuring the energy of spectator nucleons 
and fragments in a Zero Degree Calorimeter at beam rapidity. 
Following a procedure used in our previous publication
on the mid-rapidity charged multiplicity ($N_{ch}$) distribution, in which a 
clear increase of $\langle dN_{ch}/d\eta|_{\eta=0}\rangle$ per
participant with the number of participants was demonstrated~\cite{PXpub2001-1}, 
we calculate $\langle dE_T/d\eta|_{\eta=0}\rangle$ as a function of
centrality in upper percentile ranges of the 7.2b Au+Au interaction
cross section (see Table~\ref{T:1}). 
Figure~\ref{fig:4}a shows that $\langle 
dE_T/d\eta|_{\eta=0}\rangle$ per participant also increases with $N_{\rm
part}$, closely paralleling the rise in charged particle density
(Table~\ref{T:1}). This is better illustrated in
Fig.~\ref{fig:4}b where the ratio $\langle
dE_T/d\eta|_{\eta=0}\rangle/\langle dN_{ch}/d\eta|_{\eta=0}\rangle$
remains constant at a value of $\sim 0.8$~GeV, independent of centrality. 
Comparison to the measurements of WA98~\cite{rf:WA98} from Pb+Pb collisions 
at $\sqrt{s_{_{NN}}}=17.2$~GeV is instructive. The WA98 data for 
mid-rapidity $\langle dE_T/d\eta|_{\rm mid}\rangle$
per participant are shown in
Fig.~\ref{fig:4}a and are essentially independent of $N_{\rm part}$ for
$N_{\rm part} > 200$~\cite{caution}. 
WA98 parameterizes their data as
$dE_T/d\eta|_{\rm mid}\propto N_{\rm part}^{\alpha}$ with
$\alpha=1.08\pm0.06$ while the same parameterization for our data yields
$\alpha=1.13\pm0.05$. Fig.~\ref{fig:4} also shows that 
$\langle dE_T/d\eta|_{\eta=0}\rangle$ for central Au+Au collisions at 
$\sqrt{s_{_{NN}}}=130$~GeV is about 40\% larger than found by WA98, yet, for 
both c.m. energies, $\langle dE_T/d\eta\rangle/\langle dN_{ch}/d\eta\rangle$ 
remains constant versus centrality at roughly the same value, 
$\sim 0.8$~GeV (Fig.~\ref{fig:4}b).

   The Bjorken energy density for Pb+Pb collisions at $\sqrt{s_{_{NN}}}=17.2$~GeV 
was given by the NA49 collaboration~\cite{NA49ET}.  NA49 reported 
a value of mid-rapidity $dE_T/d\eta|_{\rm mid} =405$~GeV for the most
central 2\% of the inelastic cross section, in agreement with WA98. 
This corresponds~\cite{NA49ET} to a value of $\epsilon_{Bj}=2.9$~GeV/fm$^3$.  
A straightforward derivation of
$\epsilon_{Bj}$ from our measured $dE_T/d\eta|_{\eta=0}$ of
$578^{+26}_{-39}$~GeV for the same centrality cut, corrected to 
$dE_T/dy|_{y=0}$ by a factor of $1.19\pm 0.01$ from our HIJING MC, and 
taking $\pi R^2=148$ fm$^2$ (i.e. $R=1.18{\rm fm} A^{1/3}$) gives
$\epsilon_{Bj}=4.6$~GeV/fm$^3$, an increase of 60\%  
over the NA49 value. 

In conclusion, the mid-rapidity transverse energy density for central Au+Au 
collisions, and likely the spatial energy density, is at least 1.6 times 
larger at $\sqrt{s_{_{NN}}}=130$~GeV (RHIC) than at $\sqrt{s_{_{NN}}}=17.2$~GeV (CERN). 
The variation of the $E_T$ density per participant with centrality is very 
similar to the 
previously reported dependence of charged multiplicity density per 
participant at RHIC energies. These results, together with the observed 
constancy of $\langle E_T\rangle/\langle N_{ch}\rangle$ 
at a value $\sim 0.8$~GeV, 
indicate that the 
additional energy density at RHIC energies is achieved mainly by an increase 
in particle production rather than by an increase in transverse energy per 
particle. 


 
We thank the staff of the RHIC project, Collider-Accelerator, and Physics
Departments at BNL and the staff of PHENIX participating institutions for
their vital contributions.  We acknowledge support from the Department of
Energy and NSF (U.S.A.), Monbu-sho and STA (Japan), RAS, RMAE, and RMS
(Russia), BMBF and DAAD (Germany), FRN, NFR, and the Wallenberg Foundation
(Sweden), MIST and NSERC (Canada), CNPq and FAPESP (Brazil), IN2P3/CNRS
(France), DAE (India), KRF and KOSEF (Korea), and the US-Israel Binational
Science Foundation.



\vspace{4.0cm}

\vspace{-0.8cm}

\newpage

\vspace{0.5cm}
\begin{table} 

\caption[]{Average transverse energy density vs. centrality.    
The statistical errors are negligible. Errors on    
$\langle dE_T/d\eta|_{\eta=0}\rangle$ are the $N_{\rm part}$-dependent 
systematic errors from 
the uncertainty of the BBC cross section~\cite{PXpub2001-1} such that 
all points move together. There is an additional overall 
($N_{\rm part}$-independent) systematic uncertainty of $\pm 4.5$\%.}
\begin{tabular}[]{cccc}
Centrality &$\langle dE_T/d\eta|_{\eta=0}\rangle$~(GeV)&$\langle dN_{ch}/d\eta|_{\eta=0}\rangle$~\cite{PXpub2001-1}&$\langle N_{\rm part}\rangle$~\cite{PXpub2001-1}\\ \hline
 0 -  5\% &$503\pm2$ & 622 $\pm$ 41 & 347 $\pm$ 10  \\
 5 - 10\% &$409\pm4$ & 498 $\pm$ 31 & 293 $\pm$ 9   \\
10 - 15\% &$340\pm5$ & 413 $\pm$ 25 & 248 $\pm$ 8    \\
15 - 20\% &$283\pm7$ &344 $\pm$ 21 & 211 $\pm$ 7   \\
20 - 25\% &$233\pm7$ &287 $\pm$ 18 & 177 $\pm$ 7   \\
25 - 30\% &$191\pm8$ &235 $\pm$ 16 & 146 $\pm$ 6   \\
30 - 35\% &$154\pm8$ &188 $\pm$ 14 & 122 $\pm$ 5   \\
35 - 40\% &$123\pm7$ &147 $\pm$ 12 & 99  $\pm$ 5   \\
40 - 45\% &$98\pm7$ &115 $\pm$ 11 & 82  $\pm$ 5   \\
45 - 50\% &$76\pm6$ &89  $\pm$ 9  & 68  $\pm$ 4   \\
\end{tabular}
\label{T:1}
\end{table}

\end{multicols}


\begin{references}
\bibitem[*]{Deceased}Deceased

\bibitem{PHENIX1} PHENIX Collaboration, D.~P.~Morrison, {\it et al.}, 
   \Journal{\NPA} {638}{565c}{1998}.

\bibitem{QMproc} e.g. see {\sl Proc. Quark Matter 1984}, ed. K.~Kajantie 
   (Springer, Berlin 1985); {\sl Proc. Quark Matter 1987}, eds. H.~Satz, 
   H.~J.~Specht, R.~Stock, \Journal{\ZPC}{38}{1-370}{1988}. 

\bibitem{Bj83} J.~D.~Bjorken, \Journal{\PRD} {27}{140}{1983}.

\bibitem{NA34} Helios Collaboration, T.~\AA kesson, {\it et al.},  
   \Journal{\ZPC} {38}{383, 397}{1988}. 

\bibitem{PXpub2001-1} PHENIX Collaboration, 
K.~Adcox, {\it et al.}, 
\Journal{\PRL} {86}{3500}{2001}. 

\bibitem{rf:calor} E.~Kistenev et.al., {\sl 
   Proc. 5th Int. Conf. on Calorimetry in HEP}, World Scientific (1994), 
pp. 211-223; G.~David {\it et al.}, {IEEE Trans. Nucl. Sci.}\ {\bf 45},\ 692, 
705 (1998). 

\bibitem{photon} 
The PbSc EMCal measures $\sim$GeV photons 2.1\% higher than
electrons of the same energy. 
This effect is corrected in Fig.~\ref{fig:1}b. 


\bibitem{explain}
    $k=1.17=(1-0.24[inflow])/(0.79[response]\times
(1-0.18[losses]))$

\bibitem{GEANT} GEANT 3.2.1, CERN program library. 

\bibitem{HIJING} X.~N.~Wang and M.~Gyulassy, \Journal{\PRD}{44}{3501}{1991},  
version 1.35.

\bibitem{E802000} E802 Collaboration, T.~Abbott, {\it et al.}, 
\Journal{\PRC} {63}{064602}{2001}. 

\bibitem{rf:WA98} 
WA98 Collaboration, 
M.~M.~Aggarwal, {\it et al.},
   Eur. Phys J. {\bf C18},\ 651\ (2001). 

\bibitem{NA49ET} NA49 Collaboration, T.~Alber, {\it et al.}, 
   \Journal{\PRL}{75}{3814}{1995}. (The quoted $\epsilon_{Bj}=3.2$~GeV/fm$^3$ 
was divided by 1.10 to remove the enhancement factor 
applied by those authors for ``head-on'' collisions).

\bibitem{caution} $E_T$ is not Lorentz invariant; frame-dependent 
10--20\% effects in comparing fixed target experiments to colliders are 
ignored in the present discussion. 

\end{references}
\end{document}